\documentclass[twocolumn]{openjournal}

\usepackage{orcidlink}
\usepackage[T1]{fontenc}
\usepackage{ae,aecompl}

\usepackage{graphicx}	% Including figures
\usepackage{amsmath}	% Advanced maths commands
\usepackage{amssymb}	% Extra maths symbols
\usepackage{natbib}

\usepackage{hyperref}
\hypersetup{
	colorlinks=true,
	urlcolor=blue,    % color of external links
	linkcolor=black,  % color of toc, list of figs etc.
	citecolor=blue,   % color of links to bibliography
}

\defcitealias{Pittordis_2018}{PS18}
\defcitealias{Pittordis_2019}{PS19}
\defcitealias{Badry_2021}{ERH}
\defcitealias{Banik_2018_Centauri}{BZ18}

\hypersetup{citecolor=blue, % color for \cite{...} links
            linkcolor=red, % color for \ref{...} links
            menucolor=blue, % color for Acrobat menu buttons
            urlcolor=blue}  % color for \url{...} links

\newcommand{\cm}{{~\rm cm}}

\newcommand{\km}{{~\rm km}}
\newcommand{\s}{{~\rm s}}

\newcommand{\g}{{~\rm g}}

\newcommand{\K}{{~\rm K}}
\newcommand{\erg}{{~\rm erg}}

\newcommand{\days}{{~\rm days}}

\begin{document}

\title{More luminous red novae that require jets}
%\correspondingauthor{Ealeal Bear, Noam Soker}
%\email{soker@physics.technion.ac.il}

%\author[0000-0003-0375-8987]{Noam Soker}
\author{Noam Soker\,\orcidlink{0000-0003-0375-8987}} 
\affiliation{Department of Physics, Technion, Haifa, 3200003, Israel;  soker@physics.technion.ac.il}

\begin{abstract}
I study two intermediate luminosity optical transients (ILOTs)  classified as luminous red novae (LRNe) and argue that their modeling with a common envelope evolution (CEE) without jets encounters challenges. LRNe are ILOTs powered by violent binary interaction. Although popular in the literature is to assume a CEE is the cause of LRNe, I here repeat an old claim that many LRNe are powered by grazing envelope evolution (GEE) events; the GEE might end in a CEE or a detached binary system. I find that the LRN AT 2021biy might have continued to experience mass ejection episodes after its eruption and, therefore, might not suffered a full CEE during the outburst. This adds to an earlier finding that a jet-less model does not account for some of its properties. I find that a suggested jet-less CEE model for the LRN AT 2019zhd does not reproduce its photosphere radius evolution. These results that challenge jet-less models of two LRNe strengthen a previous claim that jets play major roles in powering ILOTs and shaping their ejecta and that in many LRNe, the more compact companion launches the jets during a GEE. 
\end{abstract}


% ==========================================================
\section{Introduction}
\label{sec:intro}
% ==========================================================

I adopt the theoretical definition of intermediate luminosity optical transients (ILOTs) to be all gravitationally-powered transient events in binary (or triple) stellar systems, having peak luminosities in the range of somewhat below classical novae to about typical supernova luminosities (for earlier usage of the term ILOT see, e.g., \citealt{Berger2009, KashiSoker2016, MuthukrishnaetalM2019}; there are studies that do not use the term ILOT at all, e.g., \citealt{Jencsonetal2019}). A similar, but not identical, term is gap transients, a definition that is based on observational properties. 
ILOTs, or gap transients, form a heterogeneous group with diverse time scales, luminosities, and shapes of lightcurves (a partial list of papers include, e.g.,  \citealt{Mouldetal1990, Rau2007, Ofek2008, Masonetal2010, Kasliwal2011, Tylendaetal2011, Tylendaetal2013, Kasliwal2013, Tylendaetal2015, Kaminskietal2018, Pastorelloetal2018, BoianGroh2019, Caietal2019, Jencsonetal2019, PastorelloMasonetal2019, Banerjeeetal2020, Stritzingeretal2020b, Blagorodnovaetal2021, Pastorelloetal2021, Bondetal2022,
Caietal2022a, Caietal2022b, Wadhwaetal2022, Karambelkaretal2023, Pastorelloetal2023}). 

One subclass of ILOTs is luminous red novae (LRNe). There is an observational definition of LRNe, e.g., \cite{Caietal2022a}, as those ILOTs that have a slow and long luminosity rise before the outburst, followed by two luminosity peaks; the second peak might be a plateau. LRN early-time optical spectra are similar to those of type IIn supernovae with a blue continuum with superposed narrow hydrogen emission lines. A theoretical common definition of LRNe is of ILOTs, fainter than typical supernovae, that are powered by full merger, including common envelope evolution (CEE; e.g., \citealt{Ivanova2017}), whether the companion survives or not the CEE. I note that there is no consensus on the exact definitions of these terms and the subclasses of ILOTs, e.g., \cite{KashiSoker2016} versus  \cite{PastorelloFraser2019}, \cite{PastorelloMasonetal2019}, and  \citealt{Caietal2022b}). In this study, I claim that there might be no one-to-one correspondence between the observational classification of LRNe and events experiencing a full CEE merger.
Instead, many LRNe might experience the grazing envelope evolution (GEE) during their outburst. The GEE might end with a CEE or with a detached binary system.   

One of the motivations for considering the jet-powering of ILOTs is the bipolar morphologies of spatially resolved ILOTs, some with point symmetry.  Such is the bipolar ejecta of the nineteenth-century Great eruption of Eta Carinae, called the Homunculus (e.g., \citealt{DavidsonHumphreys1997}). The Great Eruption was a luminous blue variable major eruption, a subclass of ILOTs (see Section \ref{sec:AT2021biy}). Observations show bipolar morphologies in the ILOTs V4332~Sgr \citep{Kaminskietal2018}, Nova~1670 (\citealt{Sharaetal1985, Kaminskietal2020Nova1670, Kaminskietal2021CKVul}), and V838 Mon (\citealt{Chesneauetal2014, Kaminskietal2021, Mobeenetal2021,Mobeenetal2023}). The point-symmetric morphological features of some ILOTs, e.g., the Homunculus of Eta Carinae \citep{Steffenetal2014} and  Nova~1670 (CK~Vulpeculae; e.g., \citealt{Kaminski2024}),  are key elements in claiming jet powering of ILOTs: while equatorial mass ejection during binary interaction can, in principle, explain bipolar structure, this process by itself cannot explain point symmetry. Jets are most likely to shape point-symmetric structures.  

Processes that release gravitational energy in binary interaction and can power ILOTs include mass transfer via an accretion disk where the mass-accreting (more compact) companion launches jets (e.g., \citealt{Kashietal2010, Soker2016GEE, Soker2020ILOTjets, SokerKashi2016TwoI, Kashi2018Galax}), and/or the system ejecta mass in and near the equatorial plane (e.g., \citealt{Pejchaetal2017, HubovaPejcha2019}) or the merger of two stars. The latter includes the onset of a CEE, whether the companion survives or not. Extreme cases are the spiraling-in of a neutron star or a black hole inside the envelope of a red supergiant. These events are expected to be as bright as luminous core-collapse supernovae (CCSNe) and are termed common envelope jets supernovae (CEJSNe; e.g., \citealt{SokerGilkis2018, Gilkisetal2019, GrichenerSoker2019, YalinewichMatzner2019, Schreieretal2021}). 
CCSNe, which are most likely also powered by jets (e.g., \citealt{Soker2024Rev} for a recent review), and CEJSNe, are not grouped under ILOTs despite being gravitationally-power through jets because most of the gravitational energy in these systems is carried by neutrinos rather than jets or the ejecta in general.  

In ILOTs that experience CEE, the compact companion is a star, main sequence, or slightly involved (e.g., \citealt{Tylendaetal2011, Ivanovaetal2013a, Nandezetal2014, Kaminskietal2015, Pejchaetal2016a, Pejchaetal2016b, Soker2016GEE, Blagorodnovaetal2017, MacLeodetal2017, MacLeodetal2018,  Segevetal2019, Howittetal2020, MacLeodLoeb2020, Qianetal2020, Schrderetal2020, Blagorodnovaetal2021, Addison2022, MatsumotoMetzger2022, Zhuetal2023}) or a sub-stellar (i.e., a brown dwarf or a planet) companion (e.g., \citealt{RetterMarom2003, Retteretal2006, Metzgeretal2012, Yamazakietal2017, Kashietal2019Galax, Gurevichetal2022, Deetal2023, Oconnoretal2023, Soker2023Planet}).  

Other energy sources to power the radiation of an ILOT in a CEE are the hot ejected gas (e.g., \citealt{ChenIvanova2024}), which the spiraling-in process heats, and recombination energy (e.g.,  \citealt{Ivanovaetal2013a, Howittetal2020, MatsumotoMetzger2022}). 
In \cite{Soker2023Bright}, I compared these two energy sources with jet powering and concluded that only jets could power rapidly rising lightcurves of bright ILOTs. In general, the collision of wide jets with a slower expanding ejecta efficiently channels the kinetic energy of the jets to radiation  \citep{Soker2020ILOTjets}; more efficient than equatorial ejecta collision with a shell (e.g., \citealt{Pejchaetal2016a, Pejchaetal2016b}) and a shell collision with a slow equatorial outflow (e.g., \citealt{MetzgerPejcha2017, AndrewsSmith2018, KurfurstKrticka2019}) proposed by these studies as an energy source of radiation.  In Section \ref{sec:AT2019zhd}, I examine the jet-less model that \cite{ChenIvanova2024} proposed that is based on the ejected hot gas and recombination energy. I find that it does not reproduce all the observed properties of LRN AT 2019zhd. 

Earlier studies of jet-powering focused on bumps of short-period increased luminosity, e.g., \cite{SokerKaplan2021RAA} studied the ILOT SNhunt120 (observations by \citealt{Stritzingeretal2020a}) and the LRN AT 2014ej (observations by \citealt{Stritzingeretal2020a}). In section \ref{sec:AT2021biy} I consider bumps in the lightcurve of AT 2021biy of decreased luminosity while the radius increases. I argue that these bumps suggest a continues binary interaction that might imply that no CEE took place in this outburst. I also find that the ejecta is likely to be bipolar. 
I summarize this short study in Section \ref{sec:Summary}. 

% ===========================================
\section{Challenging a jet-less model of AT 2019zhd}
\label{sec:AT2019zhd}
% ===========================================

This section addresses the dispute between models of  LRNe that attribute significant role to jets (see Section \ref{sec:intro}), and those that are not, i.e., jet-less models. 

\cite{ChenIvanova2024} construct a jet-less CEE LRN model, where the emission results from recombination and cooling of hot gas ejected by the CEE. They fit their parameters to (almost) exactly fit the lightcurve of the LRN AT 2019zhd as reported by \cite{Pastorelloetal2021}. 
 
\begin{figure*}[t]
	\centering
%	\hspace*{-2cm} 
	% [trim=left bottom right top, clip]{file}
%	\hspace{1cm}
\includegraphics[trim=0.15cm 21.2cm 0.0cm 0.0cm ,clip, scale=0.86]{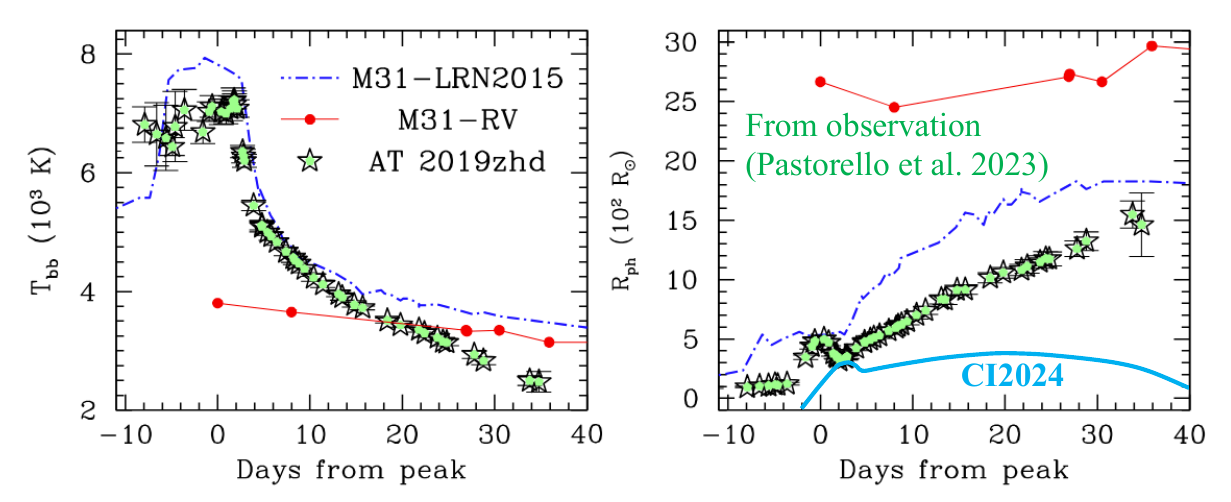} 
\caption{The evolution of the black-body temperature (left panel) and photospheric radius (right) panel that \cite{Pastorelloetal2021} calculated from their observations under the assumption of spherically symmetric ejecta. 
I added the light-blue line on the right panel, which is the radius of optical depth $\tau_{\rm R}=1$ from the jet-less model that \cite{ChenIvanova2024} built for LRN AT 2019zhd. 
Their model fails to reproduce the radius. 
}
\label{fig:AT2019zhd}
\end{figure*}

\cite{Pastorelloetal2021} calculated the blackbody temperature and radius of the photosphere of AT 2019zhd; I present two panels from their paper in Figure \ref{fig:AT2019zhd}. \cite{ChenIvanova2024} do not present the blackbody temperature nor the radius of their model. They do present the optical depth as a function of radius and time. I take their value of $\tau(r,t)=1$ to approximately give the radius of the photosphere and added the line $R(\tau=1,t)$ on the right panel of Figure \ref{fig:AT2019zhd} by a light-blue line marked CI2024.  
% FFFFFFFFFFFFFFFFFFFFFFFFFFFFFFFFFFFFFFFFFFF  
%FFFFFFFFFFFFFFFFFFFFFFFFFFFFFFFFFFFFFFFFFF

From the two lines for AT 2019zhd on the right panel of Figure \ref{fig:AT2019zhd}, I conclude that the model of \cite{ChenIvanova2024} does not reproduce the photospheric radius that \cite{Pastorelloetal2021} calculated from their observations of AT 2019zhd, not even qualitatively. \cite{ChenIvanova2024} writes also that the time evolution of the gas at $T_g= 5000 \K$ resembles the shape of the light-curve. However, as the left panel of Figure \ref{fig:AT2019zhd} shows, the blackbody temperature that \cite{Pastorelloetal2021} calculate for AT 2019zhd varies from about $7000 \K$ near maximum light to $2400 \K$ at $t=34$d. 

The jet-less model of \cite{ChenIvanova2024} and the calculations of \cite{Pastorelloetal2021}  do not agree with each other. Either one of them is wrong, or both miss the real behavior because they assume spherically symmetric ejecta and do not acknowledge the primary role that jets might play. I tend to accept the calculations of \cite{Pastorelloetal2021} as being approximately correct, but expect non-spherical effects to play significant roles (see Section \ref{sec:intro} for observational indications of bipolar ILOTs). In any case, the jet-less model that \cite{ChenIvanova2024} constructed seems not to explain the behavior of the LRN AT 2019zhd. The jets are probably needed. 
   
% ===========================================
\section{The dips in the light-curve of AT 2021biy}
\label{sec:AT2021biy}
% ===========================================

Many studies attribute the bipolar morphology of many planetary nebulae to jets, whether there are direct signatures of jets, e.g., MyCn 18 (e.g., \citealt{Bryceetal1997,  Oconnoretal2000}), or indirect, e.g., NGC 6302  (e.g., \citealt{Balicketal2023}) and NGC 7027 (e.g., \citealt{MoragaBaezetal2023}). The same holds for many bipolar symbiotic nebulae, e.g., R Aquarii, which was shaped by precessing jets (e.g., \citealt{Santamariaetal2024}), and pre-planetary nebulae (e.g., \citealt{Sahai2020}), e.g., M 1-92 (e.g., \citealt{Alcoleaetal2022}). 
Based on bipolar planetary nebulae, some past studies attributed the bipolar structure of the ejecta of the nineteenth-century Great Eruption of Eta Carinae (about 1837-1856; \citealt{Humphreysetal1999}), i.e., the  Homunculus, to powering by jets  (e.g., \citealt{Soker2001, KashiSoker2010}).
The Great Eruption of Eta Carinae was a luminous blue variable major eruption (outburst) that belongs to a subclass of ILOTs. 
I argued before \citep{Soker2023Bright} that this claim of shaping by jets applies to the Great Eruption binary scenario models that involve no CEE (e.g., \citealt{Soker2001, Soker2007}) and to the Great Eruption triple star scenarios where two of the three stars entered a CEE (e.g., \citealt{LivioPringle1998, PortegiesZwartvandenHeuvel2016, Hirai2021}). The difference is that in the non-CEE scenario, the binary system interaction might repeat itself with several peaks in the same event or repeat in consecutive events. The non-repeating nature is one of the difficulties with the triple-star scenarios of the Great Eruption \citep{PortegiesZwartvandenHeuvel2016, Hirai2021}. Specifically, the problems with the triple-star merger scenarios are as follows. 
($i$) There was a second outburst in 1890-1895 (the Lesser Eruption; \citealt{Humphreysetal1999}) that, according to the triple-star scenario,  requires another merger. Namely, a system of four stars. ($ii$) The Homunculus and the present binary system share the same equatorial plane (e.g., \citealt{Maduraetal2012}). This is difficult to explain in a non-coplanar triple-stellar system; merger scenarios require an unstable triple system likely not to be coplanar. ($iii$) The scenario proposed by \cite{Hirai2021} predicts the presence of a dense gas in the equatorial plane; this is not observed in the Homunculus. For these, I consider it unnecessary and problematic to apply a triple-star or a quadruple-star model to the two nineteenth-century eruptions of Eta Carinae. 
In the binary model for the Great Eruption of Eta Carinae, the secondary star grazed the primary stellar envelope during periastron passages. The secondary star accreted mass via an accretion disk that launched the jets that powered the event and shaped the Homunculus (e.g., \citealt{KashiSoker2010}). 

With the dispute between a one-time CEE event (e.g., \citealt{Ivanova2017}) and repeating jet-launching episodes, likely in a GEE (e.g., \citealt{Soker2020Galax}), I examine the properties of the ILOT AT2021biy that \cite{Caietal2022a} studied in detail.  
AT2021biy presents the longest plateau among LRNe, with a
duration of 210 days. \cite{Caietal2022a} find the progenitor to be a yellow supergiant of mass $\approx 20 M_\odot$, a luminosity and radius at explosion of $L_\ast \simeq 10^5 L_\odot$ and $R_\ast \simeq 300 R_\odot$. 
\cite{Caietal2022a} tried to fit the light-curve with the jet-less recombination model of \cite{MatsumotoMetzger2022}: From the plateau duration of $t_{\rm pl}=210 \days$ and the expansion velocity of $\bar v_{\rm E} =430 \km \s^{-1}$ that they deduced from the H$\alpha$ line they estimated the ejected mass to be $M_{\rm ej} \simeq 9.9 M_\odot$. With these parameters, they found the jet-less model of \cite{MatsumotoMetzger2022} to give a plateau luminosity of $\approx 3 \times 10^{39} \erg \s^{-1}$, which is an order of magnitude below the observed plateau luminosity $L_{\rm pl} \simeq 5 \times 10^{40} \erg \s^{-1}$. Therefore, \cite{Caietal2022a} claimed an extra energy source powered the plateau. I attribute the extra energy to jets, or more accurately, most of the energy to jets.    
 
\begin{figure*}[t]
	\centering
%	\hspace*{-2cm} 
	% [trim=left bottom right top, clip]{file}
%	\hspace{1cm}
\includegraphics[trim=0.0cm 9.2cm 6.0cm 0.0cm ,clip, scale=0.96]{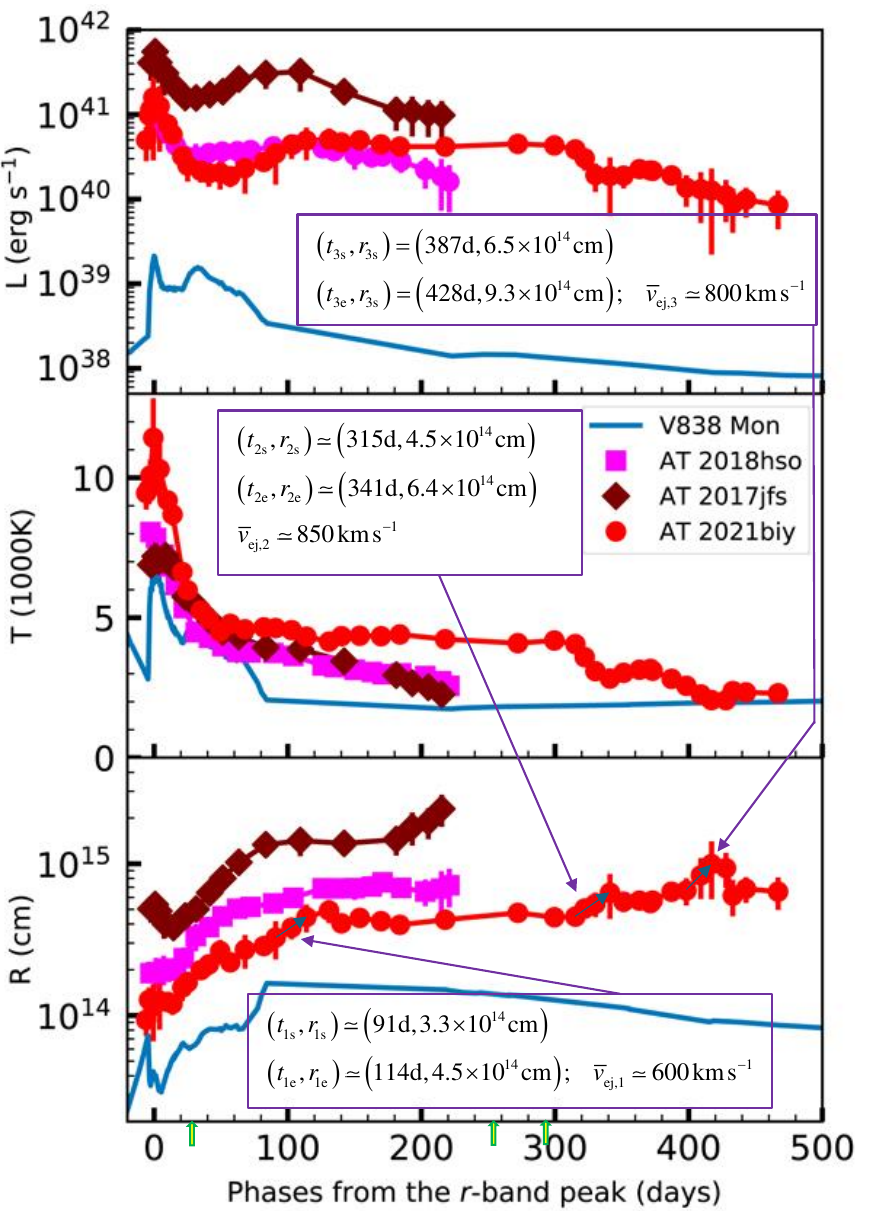} 
\caption{A figure adapted from \cite{Caietal2022a} of the bolometric lightcurves, blackbody temperature evolution, and blackbody (photosphere) radius evolution of four ILOTs. I added the estimated starting and ending times and radii (three insets) of three rapid radius increase periods of AT2021biy, as indicated by the three arrows in the lower panel. The insets also indicate the expansion velocity of each rise, namely $v_{\rm ej}=(r_{\rm e}- r_{\rm s})/ te-t_{\rm s})$ for each of the three rapid rises. The green arrows indicate the estimated ejection time of the material calculated by $t_{\rm ej} =t_{\rm s} - r_{\rm s}/ v_{\rm ej}$. I expect that there were other jet-launching episodes that did not obscure the equatorial photosphere. 
}
\label{fig:AT2021biy}
\end{figure*}

Figure \ref{fig:AT2021biy} from \cite{Caietal2022a} presents the luminosity, blackbody temperature, and photosphere radius of four ILOTs. I refer to AT 2021biy. I examine three radius rises as marked on the lower panel. In the insets, I give the start and end times of the three time periods and the radii at these times. I take the times and radii from table A.2 of \cite{Caietal2022a}. From the start and end points, I calculate the average velocity of the increasing radius. The radii of these points are highly uncertain, and therefore, so is the velocity I calculate. Assuming that this velocity is the velocity of a mass, I calculated the time the central stellar system ejected the mass. I mark these times with the green-double-lined arrow on the time axis. 
% FFFFFFFFFFFFFFFFFFFFFFFFFFFFFFFFFFFFFFFFFFF  
%FFFFFFFFFFFFFFFFFFFFFFFFFFFFFFFFFFFFFFFFFF

A rapid rise in radius can result from a mass that the stellar system ejected late and catches up with the photosphere. What is different in the present case for the two late radius rises is that decreases in luminosity rather than increases accompany these increases in radius. A very moderate rise in luminosity accompanies the first radius rise. In the case of shell collision, the expectation is an increase in luminosity because kinetic energy is channeled to thermal energy that is partially radiated away. The two late radius rises end with a temperature where most of the gas is recombined. Therefore, it seems that the shell that catches up with the photosphere is mainly neutral, and the collision with the rarefied photospheric gas does not heat it much. 

The flow structure is likely to be non-spherical. For example, the main photosphere might be from an equatorial mass ejection before the rapid radius rise. The late ejecta might be then a wide bipolar outflow, say of a half opening angle of $\alpha_{\rm j} \sim 45^\circ - 60 ^\circ$. If we observe this LRN along the polar direction, the cold polar ejecta obscures the equatorial photosphere for several weeks; this explains a cool expanding photosphere. In that case, there might be more mass-ejection episodes of lower mass, those that do not obscure the equatorial gas. The possible parameter space of these late bipolar mass-ejection episodes is large. It deserves a study by itself. 

The mass in the latest ejecta that caused the third rapid radius rise can be estimated as follows. The blackbody temperature of the gas in the last rise at maximum radius $r_{\rm 3e} \simeq  10^{15} \cm$, at $t \simeq 410-430 \days$, is $T_{\rm BB,3e} \simeq 2000 \K$. At that low temperature, molecular opacity dominates, and the opacity is $\kappa \simeq 0.01 \cm^2 \g^{-1}$ (e.g., \citealt{FergusonAlexander2005, Fergusonetal2007}), for a density of $\simeq 10^{-11} \g \cm^{-1}$. 
The required mass to have an optical depth of 1 at that radius is 
\begin{equation}
M_3 \simeq 0.63 \left( \frac{r_{\rm 3e}}{10^{15} \cm} \right)^2 \left( \frac{\kappa}{0.01 \cm^2 \g^{-1}} \right)^{-1} 
\left( \frac{ \Omega}{4 \pi} \right) M_\odot .
\label{eq:M3}
\end{equation} 
The total mass is smaller if the mass is in a wide polar outflow rather than a complete shell over a solid angle of $\Omega=4 \pi4$. For a half opening angle of $60^\circ$ ($45^\circ$; measured from the polar axis), the ejected mass is 
$M_3 \simeq 0.3 M_\odot$ ($M_3 \simeq 0.2 M_\odot$).  

If the same approach is applied to the second rise, the required ejected mass is too high. The temperature at the largest radius in the second rise, $r_{2e} \simeq 6.4 \times 10^{14} \cm$, is $T_{\rm BB,2e} \simeq 3000 \K$ where the opacity is only $\kappa \simeq 3 \times 10^{-4} \cm^2 \g^{-1}$ (e.g., \citealt{FergusonAlexander2005, Fergusonetal2007}).
This implies, for a spherical shell by equation (\ref{eq:M3}), an ejected shell mass of $M_2 \simeq 8.6 M_\odot$.  This is too large for the inferred progenitor system. A more likely explanation is that the ejecta is wide polar ejecta (to two pools). This reduces the required ejected mass from two effects. The first effect is that $\Omega < 4 \pi$. The second is that the polar ejecta does not cover the entire photosphere of the equatorial ejecta. In that case, it is possible that the infrared temperature of $T_{\rm BB,2e} \simeq 3000 \K$  is a combination of two photospheres: One of the equatorial ejecta at $\approx 4000 \K$, and one is of the polar ejecta at $\approx 2000 \K$. At both these temperatures, the opacity is higher than at $\approx 3000 \K$, which implies a lower required mass. This implies that the ejected polar mass that obscures some of the equatorial photosphere is $M_2 \approx 0.1M_\odot$. 

In the case of AT 2021biy, my suggestion is that the fast polar outflow recombines before it covers a large area, and, therefore, when it expands enough to start obscuring the equatorial photosphere, it is already cold at $\approx 2000 \K$. 
The polar and equatorial outflows collide on their overlapping solid angles, and the polar outflow further heats the equatorial outflow and prolongs the plateau. As said, I expect that there have been earlier jet-launching episodes that did not obscure the equatorial photosphere but heated it up. 

Further detailed research is needed on the structure of two photospheres at different temperatures in a bipolar outflow: a denser and slower thick equatorial outflow and a faster eruptive bipolar outflow. 

My conclusion from this short discussion is that the lightcurve of LRN AT2021biy, that, as noticed by \cite{Caietal2022a}, the jet-less model of \cite{MatsumotoMetzger2022} cannot explain, was powered by late jet-launching episodes. This implies, in turn, that the interaction of the main outburst was more likely to have been a GEE than a CEE. Whether the binary system eventually entered a CEE or survived as a detached binary system is for future observations to determine. In the latter case, this system can experience another ILOT event, much as the Lesser Eruption of Eta Carinae at the end of the nineteenth century.

% ===========================================
\section{Summary}
\label{sec:Summary}
% ===========================================

This study is motivated by two recent ILOTs classified as LRNe, AT 2019zhd  and AT 2019zhd, which jet-less models based on recombination energy have difficulty explaining. In Section \ref{sec:AT2019zhd}, I found that the jet-less model of  \cite{ChenIvanova2024} does not reproduce the radius evolution of LRN AT 2019zhd as \cite{Pastorelloetal2021} calculated from their observations. As noticed by \cite{Caietal2022a}, the jet-less model of \cite{MatsumotoMetzger2022} cannot reproduce both the luminosity and duration of LRN AT 2021biy. In section \ref{sec:AT2021biy}, I further argued that the late lightcurve of AT 2021biy requires late mass-ejection episodes. As earlier studies suggested (see section \ref{sec:intro}), a late activity that powers more energy to an ILOT and which also accounts for bumps in the lightcurve might be the launching of jets by the more compact companion in a binary system. I suggest this late binary activity to the LRN AT 2019zhd. This implies that the binary system did not enter a CEE at the main outburst. 

For both these LRNe, I take the view that the binary interaction during the main outburst involved GEE rather than CEE. 
In \cite{Soker2016GEE}, the last sentence of the Abstract reads: ``Some future ILOTs of giant stars might be better explained by the GEE than by merger and CEE without jets.'' the progenitor of AT 2021biy was a yellow giant ($R_\ast \simeq 300 R_\odot$). I attribute the LRN AT 2021biy to a GEE event. It might have ended in a final CEE, or the two stars might have stayed detached. 
I repeat my earlier claim that many ILOTs are powered by a GEE event; it might end with a CEE or a detached binary system.   
There are no observational indications for the progenitor of AT 2019zhd, but I expect jets to have powered this ILOT as well. 

This study adds to the accumulating shreds of evidence, observationally and theoretically, that jets play major roles in powering most ILOTs, including all subclasses of LRNe and luminous blue variable major eruptions, whether the system enters a CEE during the outburst or experiencing a GEE. Note that in some LRNe, the binary interaction might be with a sub-stellar companion.

% ===================================================
\section*{Acknowledgements}
% ===================================================
A grant from the Pazy Research Foundation supported this research.

%%%%%%%%%%%%%%%%%%%%%%%%%%%
% ===================================================
%\section*{Data availability}
% ===================================================
%\textbf{Data availability}
% The data underlying this article will be shared on reasonable request to the corresponding author. 
%%%%%%%%%%%%%%%%%%%%%%%%%%%

%\label{lastpage}
\end{document}